\documentclass[%
 aip,
 amsmath,amssymb,
 reprint,%
]{revtex4-1}

\usepackage{graphicx}
\usepackage{dcolumn}
\usepackage{bm}

\usepackage[utf8]{inputenc}
\usepackage[T1]{fontenc}
\usepackage{mathptmx}

\newcommand{\nc}{\newcommand}
\nc{\nn}{\nonumber}
\def\e{\mathcal{E}}
\nc{\XYZ}[1]{#1} 
\begin{document}

\preprint{AIP/123-QED}

\title{Recent developments in plasmon-assisted photocatalysis - a personal perspective}

\author{Yonatan Sivan}
\affiliation{School of Electrical and Computer Engineering, Ben-Gurion University, Beer-Sheva, 8410501, Israel}
\email{sivanyon@bgu.ac.il.}
\author{Yonatan Dubi}
\affiliation{Department of Chemistry, Ben-Gurion University of the Negev, Beer-Sheva, 8410501, Israel}

\date{\today}

\begin{abstract}
It has been known for many years that metallic nanoparticles can catalyse various chemical reactions, both in the dark and under illumination, through different mechanisms. In the last decade or so, many claims of plasmon-assisted ``hot'' electron driven catalysis of bond-dissociation reactions have been put forward. These claims were \XYZ{challenged} in a recent series of papers, where both the underlying theory of ``hot'' electron generation and the use of specific experimental setups to discover them in chemical reactions were examined in detail. The conclusion that arose from these works is that as long as temperature gradients exist inside the system (as for typical experimental setups) a quantification of non-thermal effects is close to impossible. Instead, a standard thermal theory was shown to be capable of explaining the experimental findings quite accurately. Here, we review the central lines of thought that led to these conclusions, from a personal perspective. We lay out the key aspects of the theory, and point to the specific caveats one must be aware of in performing photo-catalysis experiments. Finally, we provide some future directions of study.
\end{abstract}

\maketitle

Metal nanoparticles (NPs) are known to exhibit unusual optical properties, in particular, for having absorption and scattering cross-sections that exceed their physical size. This was predicted to enable a long list of potential applications in diverse fields~\cite{Giannini_chemrev}. Unfortunately, however, the strong absorption in the metal severely limits many of these applications~\cite{Khurgin_Boltasseva_MRS,Khurgin_loss_Nat_nanotech}. Since the energy absorbed in the metal gets converted to heat almost in its entirety~\cite{Dubi-Sivan,Dubi-Sivan-Faraday}, in more recent years, attention was diverted into applications that {\em exploit} the absorption, like high temperature chemistry, thermal treatment of cancer, water purification etc.~\cite{thermo-plasmonics-review,Baffou_solvothermal,refractory_plasmonics,solar_steam_apps,solar_steam_apps_2}.

Adding to these, one of the most promising such applications was predicted to be the speeding up of chemical reactions, usually referred to as plasmon-assisted photocatalysis. In particular, it was envisioned that the absorption of photons promotes the generation of a non-equilibrium (also known as non-thermal, or ``hot'') carrier distribution such that electrons in the high-energy tail of this distribution can tunnel out of the metal into high-energy orbitals of the surrounding molecules, and then catalyse the chemical reaction. This was usually referred to as the ``hot carrier mechanism'' (or ``indirect'' plasmon-assisted photocatalysis~\cite{wei2018quantum,zhang2016fundamental,seemala2019plasmon}). \XYZ{We note that the term ``hot carriers'' is somewhat unfortunate, as it implies that the carriers are in thermal equilibrium (i.e. their distribution is Fermi-like) albeit with a temperature higher than the ambient (such that the system in general is out of equilibrium). Yet, the common practice in the field refers to ``hot electrons'' as having a distribution which is different than a thermal distribution with the ambient temperature (see, e.g., the title of Ref.~[\onlinecite{Halas_Science_2018}]). Hereafter, we use the terms ``non-thermal'' and ``hot'' intermittently, as is accustomed in the literature}.

The ``hot'' electron mechanism was introduced to explain two main phenomena. First, it explained the ability of a high energy metal electron to tunnel through the Schottky barrier to an adjacent semiconductor and then be used for photodetection at frequencies lower than the gap energy of the semiconductor\cite{russell2003room,sandhu2000near,Moskovits_hot_es_water_splitting,Moskovits_photosensitization,Uriel_Schottky,Uriel_Schottky2,Valentine_hot_e_review}.  Elaborate models of this effect provided a very good quantitative agreement with experimental findings (see, e.g.,~[\onlinecite{Giulia_Nat_Comm_2018}]). Second, motivated by this first effect, the ``hot carrier mechanism'' was also employed to explain why {\em noble} metals, which are not known as particularly good catalysts, did provide significant catalysis when subjected to optical illumination. The effect was primarily explained as enabling {\em low} activation energy pathways. In some studies (e.g.~[\onlinecite{Halas_dissociation_H2_TiO2,Halas_H2_dissociation_SiO2,christopher2011visible,plasmonic_photocatalysis_Linic,Halas_Science_2018}]), it was claimed that the mere presence of ``hot'' carriers reduces the activation energy of the favourable reaction pathways, as a function of their number density (hence, as a function of the incoming light intensity). This approach was predicted to surpass the efficiency of traditional catalysis approaches~\cite{liu2018metal,chem_rev_photochemistry_2006} and to circumvent the well-known limitations associated with catalysis using high temperatures. The latter includes high energy-consumption, shortened catalyst lifetimes~\cite{thermal_shortening_catalyst_lifetime} through sintering deterioration, and especially non-selectivity which enables undesired reactions to take place, hence, to loss of yield and efficiency, see e.g., discussion in~[\onlinecite{hot_e_review_Purdue}]. This conclusion led to a rapid growth of interest in plasmon-assisted photocatalysis, mostly as a viable pathway towards cheap and efficient way to produce ``green'' fuels~\cite{hirakawa2004photoinduced,plasmonic-chemistry-Baffou,plasmonic_photocatalysis_Clavero,hot_es_review_2015_Moskovits,Valentine_hot_e_review,hou2013review,wu2017direct,brooks2018toward,zhang2017surface}.

Potentially because of the complex multi-disciplinary nature of this problem, the exact manner in which ``hot'' electrons assist the reaction rate was not supported by a quantitative, first-principle type theory, but instead, has remained at the phenomenological level. Moreover, not only was it unclear how the ``hot'' electrons mechanism related to the so-called ``direct'' mechanism~\cite{Wolf-Ertl,Petek_Nat_Phot_2017} traditionally associated with {\em ``standard''} metal catalysts such as Pd, Pt, Ru, Rh etc. (see discussions in~[\onlinecite{Y2-eppur-si-riscalda,Naldoni-tutorial-2020}]), this explanation also ignored the role of mere near-field enhancement~\cite{Petek_Nat_Phot_2017}, as well as the build up of temperature that follows the decay (thermalization) of the ``hot'' electrons~\cite{Aeschliman_e_photoemission_review}. The relative importance of these effects is particularly interesting in the context of standard catalysts (see e.g.,~[\onlinecite{Liu-Everitt-Nano-Letters-2019}]), and even in the so-called antenna-reactor systems (see, e.g.,~[\onlinecite{JACS_Xiong,Swearer_antenna_reactor,Boltasseva_LPR_2020}]) which combine a plasmonic metal (as a light harvester) and a catalytic metal (as a reaction site).

From a more general perspective, the relative importance of thermal and non-thermal effects in illuminated plasmonic NPs remained an issue under debate~\cite{culver1996temperature,leenheer2014solar,Baldi-ACS-Nano-2018,sarhan2019importance}. In this context, while in some works the importance of thermal effects was acknowledged and harnessed for useful applications~\cite{thermo-plasmonics-review,Govorov_thermoplasmonics,refractory_plasmonics}, many early (experimental as well as theoretical) studies overlooked thermal effects or concluded that they were small. In some cases, this may have originated from the conceptually difficult distinction between thermal and non-thermal effects, or from the common incorrect conception that at low illumination intensity, non-thermal effects dominate over thermal effects~\cite{Dubi-Sivan,Dubi-Sivan-Faraday}. Other studies estimated thermal effects crudely and/or employed too simplistic control experiments, most likely because of the limitations of existing reliable thermometry techniques, especially in the early stages of this line of research. \XYZ{The fact that quantum mechanical effects on the nanoscale are more attractive than ``macroscale'' thermal effects may have also offered some incentive for the downplaying of the latter. }

In a series of recent works, we have shown how to separate thermal and non-thermal effects using a simple addition to standard theoretical approaches~\cite{Dubi-Sivan,Dubi-Sivan-Faraday}. We have also shown that standard modelling and careful temperature measurements can provide a purely thermal quantitative explanation to many (although not all) reports of faster chemical reactions in the presence of illuminated metal nanoparticles~\cite{anti-Halas-Science-paper,Y2-eppur-si-riscalda,R2R,Un-Sivan-sensitivity}.

Below, we review this line of work, aiming at the non-expert audience. Notably, we do not aim to review all the vast literature on the topic (especially not all the experimental work), noting that since our first publications in the field~\cite{Dubi-Sivan-Faraday,anti-Halas-Science-paper}, several review papers and viewpoints have already been published~\cite{thm_hot_e_faraday_discuss_2019,dyn_hot_e_faraday_discuss_2019,Baffou-Quidant-Baldi,Khurgin-Faraday-hot-es,Khurgin-nanophotonics-2020,Jain_viewpoint,Naldoni-tutorial-2020}. Rather, we wish to take the reader through the reasoning that guided us in our studies and how it collides with many of the conclusions drawn previously in the literature. More generally, our aim in this perspective is to promote the implementation of simple, intuitive and quantitative physical arguments before employing exciting yet speculative and/or highly sophisticated ones.

{\bf Evidence for thermally-driven photo-catalysis from theory and analysis of experimental data. }
The starting point of our work was a simple question - what happens to a small piece of metal when it is continuously illuminated? This simple looking question turns out to be hard to answer. Specifically, there seems to be a clash between the n\"aive intuitive answer ``it heats up'' and the strict physical statement that since the metal is out of equilibrium, temperature is no longer well-defined and one cannot talk about heating at all. This may have been the reason that several theoretical papers tried to answer the aformentioned question by considering only how light would affect the electron distribution inside the NP (see e.g., Refs.~[\onlinecite{Manjavacas_Nordlander,Govorov_1}]) or considering heating by taking the temperature of the electrons as a fixed parameter (and guessing it rather than calculating it), and even worse, assuming that phonons do not heat up at all~\cite{Govorov_ACS_phot_2017}. It is, however, not hard to appreciate that while ignoring heat generation and heat leakage to the environment may be valid at the early stages of an ultrashort excitation, these effects cannot be ignored when studying the steady-state case.

To reconcile the apparent paradox described above, in Ref.~[\onlinecite{Dubi-Sivan}] we took a simple route of using the Boltzmann equation, which takes into account all possible energy transfer pathways in the system, and augmenting it with nothing more than the law of energy conservation in the {\em whole} system, i.e., including not only the photons and electrons, but also the lattice and the environment. \XYZ{More specifically, we demanded that at the steady-state, all the power that is pumped into the system via photon absorption must leave it via heat current to the surrounding host. Further simplification was obtained by naturally separating the electron distribution into a ``thermal'' and `non-thermal'' parts, where the thermal part is a Fermi distribution, with a temperature which is {\sl different} than the ambient temperature. These two points allowed us to calculate correctly both the electron temperature and the steady-state electron non-equilibrium distribution (including the ``hot'' electrons) under continuous wave illumination. } Thus, unlike previous studies of the electron non-equilibrium, this approach allowed us to evaluate the full non-equilibrium electron distribution, while naturally accounting for the effect of nanoparticle size and shape, as well as the thermal properties of the host on the rate of heat transfer to the environment.


Two central findings came out of this approach. The first is that, due to the fact that the electron-electron relaxation time is incredibly short in metals, almost all of the power coming from the illumination goes into heating the electrons and the phonons (which maintain almost the same steady-state temperature), and only a tiny fraction goes into skewing the electron distribution and generating ``hot'' electrons. Importantly, our calculations have shown clearly that the dominance of thermal effects over non-thermal (``hot'' carrier) effects become {\em more} significant as the illumination intensity becomes lower. This result {\em invalidates} a common claim~\cite{Govorov_ACS_phot_2017} that since the temperature rise associated with low illumination intensity is small with respect to the ambient temperature, then, ``hot'' carrier effects are dominant for low intensities. The second finding is that in spite of the first point above, the {\sl number} of ``hot'' electrons, i.e., the number of electrons with an energy excess of what they would have within a thermal-only distribution, increases by {\em many} orders of magnitude.

These two findings seem to be at odds with many claims of photocatalysis being accelerated by illuminated plasmonic nanoparticles, most specifically with some of the seminal papers in the field~\cite{Halas_dissociation_H2_TiO2,Halas_H2_dissociation_SiO2,christopher2011visible,plasmonic_photocatalysis_Linic, Halas_Science_2018}. {\XYZ Specifically, simple estimates of heating suggest much higher heating than observed in these studies (the reasons for that will be discussed below). In addition, only a 1-2 orders of magnitude rise in reaction rates was observed with respect to the reaction rate in the dark, although the $8-10$ orders of magnitude increase of the {\sl number} of ``hot'' electrons implies that a similar $8-10$ orders of magnitude enhancement should be observed in the reaction rate~\cite{Dubi-Sivan-Faraday}.}

Following these apparent discrepancies, the question arises: can experimental reports of faster reaction rates in the presence of metal NPs be explained solely based on thermal effects, assuming that the only thing that is happening in experiments is that the sample heats up? The answer to this question turns out to be yes.

It requires, however, recognizing that the {\sl measured} temperature, $T_M$, may be lower than the {\sl actual} temperature $T$ of the catalyst (this assumption will be proven correct in the following sections, as was discussed in Refs.~[\onlinecite{anti-Halas-Science-paper,Y2-eppur-si-riscalda}]). Why is this important? because, according to the standard Arrhenius theory of chemical reactions, the reaction rates obey
\begin{equation}\label{EQ:Arrhenius}
R \propto \exp \left(-\frac{\e_a}{k_B T}\right),
\end{equation}
where $\e_a$ is the activation energy, $k_B$ is the Boltzmann constant. Thus, if the {\sl measured} temperature, $T_M$, is smaller than the actual temperature $T$ of the catalyst (as shown in~\cite{Y2-eppur-si-riscalda}~\footnote{Indeed, in~\cite{Y2-eppur-si-riscalda} we have shown that $T_M$ was several tens of degrees lower than the actual temperature, and sometimes~(\cite{Halas_Science_2018,R2R}) even several hundreds of degrees lower. }), it would appear that the reaction rates overshoot the Arrhenius law - a result which thus was interpreted as enhanced reaction rate coming from ``hot'' electrons.

To demonstrate that this is a plausible claim, we rely on the simple connection between the catalyst temperature, the measured temperature and the incident illumination intensity $I_{inc}$,
\begin{equation}\label{EQ:T}
T = T_{dark} + a I_{inc} = T_M + \tilde{a} I_{inc}.
\end{equation}
where $T_{dark}$ is the temperature of the reactor when no illumination is present. The photothermal conversion coefficient $a$ depends on a number of system-specific parameters (NP size and shape, material, density and number, illumination wavelength, thermal properties of the host etc.)~\cite{thermo-plasmonics-basics,thermo-plasmonics-review,Baffou_pulsed_heat_eq_with_Kapitza,thermo-plasmonics-multi_NP}. As explained in detail in~[\onlinecite{R2R}], Eq.~(2) can be easily extended to account for the slower rate of heating occurring at relatively high illumination levels, e.g., by adding a nonlinear contribution of $I_{inc}$ to the temperature. However, this is necessary only in very few experiments, see~[\onlinecite{Y2-eppur-si-riscalda}] and~[\onlinecite[p. 270-271]{thm_hot_e_faraday_discuss_2019}].

Eqs.~(\ref{EQ:Arrhenius}) and~(\ref{EQ:T}) have only two unknowns for any given photocatalytic reaction, namely the activation energy $\e_a$ and the photo-thermal conversion coefficient $\tilde{a}$. In order to obtain these from published data, all one needs is the reaction rate {\sl in the dark} (which provides $\e_a$) and a {\sl single point} of reaction rate under illumination at a given illumination intensity (which provides $\tilde{a}$). This strikingly simple procedure is enough to reproduce essentially {\sl all the data} of Refs.~[\onlinecite{Halas_dissociation_H2_TiO2,Halas_H2_dissociation_SiO2,christopher2011visible,plasmonic_photocatalysis_Linic,Halas_Science_2018}] with remarkable accuracy\cite{anti-Halas-Science-paper,Y2-eppur-si-riscalda,R2R}. In Fig.~\ref{fig:collage} we show a collage of experimental data from Refs.~[\onlinecite{Halas_Science_2018}](a), [\onlinecite{Halas_dissociation_H2_TiO2}](b),~[\onlinecite{plasmonic_photocatalysis_Linic}](c) and~[\onlinecite{christopher2011visible}](d), along with a fit to an Arrhenius curve (solid line), demonstrating the agreement between the data and the theory described above.

{\XYZ Nevertheless, as noted in Ref.~\onlinecite{Y2-eppur-si-riscalda}, from an Arrhenius fit alone, one cannot determine {\XYZ with absolute certainty} if the illumination only changes the temperature without changing the activation energy, or contributes to both effects (a point which was later raised also by Jain~\cite{jain2019phenomenological,response_2_Jain}). Thus, in order to gain further support to the pure thermal interpretation, the fits shown in Fig.~\ref{fig:collage} were then validated with {\em independent} thermal calculations which accounted for the details of the metal NPs used (shape, size, density, ..), the host material and illumination via an effective medium theory. The excellent agreement between the values of the calculated and fitted photothermal conversion coefficient $a$ (e.g., Ref.~\onlinecite{anti-Halas-Science-paper}) indicates that one does not need to invoke any mechanism of change in the activation energy in order to explain the experimental results. }

\begin{widetext}
\begin{center}
\begin{figure}[h!]
\includegraphics[width=17.5truecm]{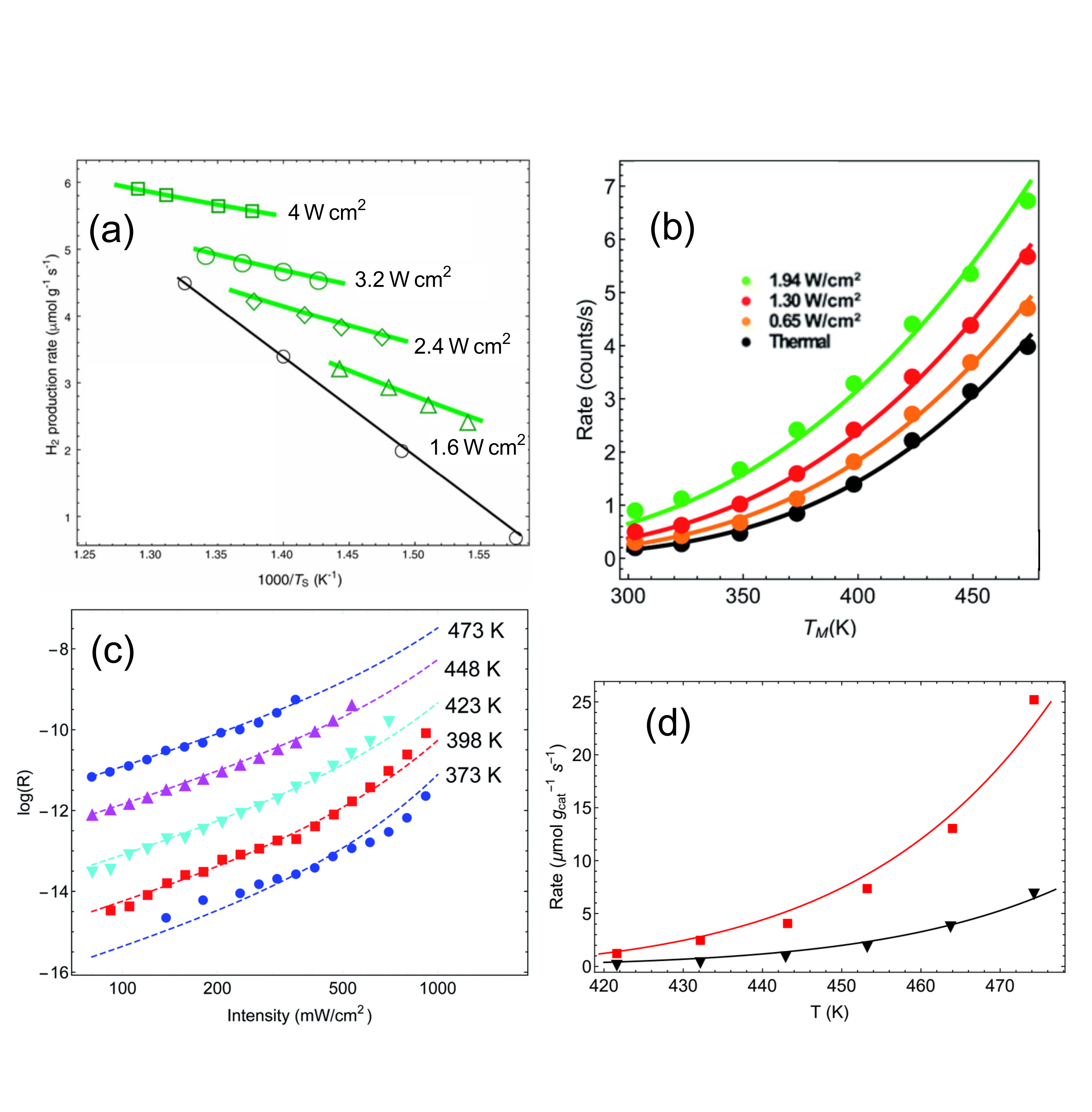}
\caption{Experimental data (symbols) and a fit (lines) using the classical Arrhenius theory of Eq.~(\ref{EQ:Arrhenius}), showing remarkable agreement between theory and experiment. Data are collected from Refs.~[\onlinecite{Halas_Science_2018}](a), [\onlinecite{Halas_dissociation_H2_TiO2}](b),~[\onlinecite{plasmonic_photocatalysis_Linic}](c) and~[\onlinecite{christopher2011visible}](d). Figures are reproduced with permission from Refs.~[\onlinecite{anti-Halas-Science-paper}](a) and~[\onlinecite{Y2-eppur-si-riscalda}](b-d). }\label{fig:collage}
\end{figure}\end{center}
\end{widetext}

\vskip 0.5truecm
{\bf On the importance of understanding temperature gradients. } The remarkable reconstruction of the experimental data with Arrhenius theory confirms that there is a substantial difference between the temperature of the catalysts ($T$) and the {\sl measured} temperature ($T_M$) in the studies analyzed in Fig.~\ref{fig:collage}. The reconstruction is also based on the assumption that one can characterize the catalysts using a single temperature. But what is this temperature and how does it relate to the reality of the experiments? These questions led us to a comprehensive inquiry into the {\em experimental} procedures of the papers we analyzed~\cite{Halas_Science_2018,Halas_dissociation_H2_TiO2,christopher2011visible,plasmonic_photocatalysis_Linic,Halas_Science_2018}. We discovered a series of flaws in the experimental setups, that may have led the authors of these papers to exaggerate the importance of ``hot'' electrons compared to thermal effects. A detailed account of proper experimental practices required to separate thermal from ``hot'' electron contributions has been given by us~\cite{Y2-eppur-si-riscalda} and by others~\cite{Liu-Everitt-Nano-Letters-2019,Liu-Everitt-Nano-research-2019,Jain_viewpoint,Baffou-Quidant-Baldi}.

Here, we wish to stress what we believe is a key issue in all experiments attempting to separate thermal from non-thermal contributions, namely, the correct account for temperature gradients. This topic has been discussed at length in~[\onlinecite{Liu-Everitt-Nano-Letters-2019,R2R}], so we wish to demonstrate its importance via an example. Consider a (rather standard) photo-catalysis experiment, where a $\sim 1$mm sample is illuminated from above and heated from below. How would one go about and isolate the non-thermal contribution to the reaction rate? A naïve approach would be to measure the reaction rate under illumination, record the surface temperature, measure the reaction rate at the recorded temperature in the dark, and subtract the latter reaction rate from the former. Such a protocol has indeed been employed on various works to extract the ``hot'' electron contribution to the reaction rate~\cite{Halas_Science_2018,Halas-Nature-Catalysis-2020}.

However, there is an essential flaw in this protocol, as it completely neglects temperature gradients~\cite{Liu-Everitt-Nano-Letters-2019}, and here is why. When the reaction rate is measured under illumination, the surface of the sample is much hotter than its bulk, because the penetration of light is rather small (typically tens of microns). Thus, effectively, the bulk of the sample will contribute to the catalysis at a reaction rate which is defined by Eq.~(\ref{EQ:Arrhenius}) with a temperature which is much lower than the surface temperature.

On the other hand, heating the sample using an external heater (i.e., in the dark), typically leads to a more uniform temperature distribution (in fact, there might be temperature gradients in the reverse direction to the illuminated case, see~[\onlinecite{Liu-Everitt-Nano-Letters-2019}] in the presence of a bottom heater; this possibility is ignored for the sake of simplicity). In these cases, when heating to the surface temperature, the entire bulk is at the surface temperature recorded in the photocatalysis experiment. Therefore, subtracting from the illuminated reaction rate the reaction rate in the dark at the surface temperature is meaningless. Similarly, the non-negligible transverse gradients (see~[\onlinecite{Un-Sivan-sensitivity}]) are likely to cause additional differences between the experiment and its control; higher-order moments of the temperature distribution may be playing a role, yet to be understood~\cite{Liu-Everitt-Nano-Letters-2019}$^,$~\footnote{Note that the transverse gradients were not accounted for in that work.}.

The only way around this problem, at least for thick samples, is to have a thermal control experiment in the dark that fully reproduces the temperature gradient under illumination. While attempts in this direction have been made~\cite{Liu-Everitt-Nano-research-2019}, in reality this is probably an impossible task, essentially because of the exponential sensitivity of the reaction rate on the temperature. The remaining alternative is to perform a thermocatalysis control in which the sample is hotter than in the photocatalysis experiment, see, e.g.,~[\onlinecite{yu2019plasmonic}]; this approach may highlight the contribution of the `hot'' electrons, but is incapable of properly quantifying it.

{\bf More systematic evaluation of the temperature distribution in plasmon-assisted photocatalysis samples. } In~[\onlinecite{Un-Sivan-sensitivity}], we generalized the discussion of the temperature calculations for plasmon-assisted photocatalysis samples. First, we validated the effective medium approach presented in~[\onlinecite{Y2-eppur-si-riscalda}], and provided a simple analytic formula for the temperature rise in characteristic samples. This formula enables an immediate evaluation of the importance of the thermal effects; it also opens the way to inclusion of fluid dynamics effects, see below. Then, we studied the dependence of the temperature distribution on the various parameters of the system. We showed that since typically these samples are designed to absorb all the incoming light, the steady-state temperature distribution shows overall {\em in}sensitivity to the illumination spectra, temporal pattern, particle density, size and shape; this result {\em contradicts} a wide range of popular claims that can be found in prior literature and verbally-presented conceptions, which usually do not follow from a proper thermal analysis. In that sense, the temperature distribution is determined by macroscopic considerations, i.e., the balance of total heat generated in the sample and the heat transfer to the surroundings (rather than by the microscopic details). This also shows that while the use of metal nanoparticles for heating was initially motivated by the localized nature of heat generation at the nano-proximity of the particles, in practice, the (steady-state) temperature distribution generated by an ensemble of NPs is not so different from that created by an external heat source.

The results of this analysis reflects some rather intuitive conclusions. First, thermal effects accumulate with the number of NPs, whereas ``hot'' carrier effects do not - for a given level of illumination, they are independent of the number of NPs. This implies that further to the local considerations presented above (see~[\onlinecite{Dubi-Sivan-Faraday}]), reactions in dense systems of NPs are more likely to be catalyzed by heating rather than by non-thermal effects.

Second, while the initial intention was to speed up chemical processes with high activation energy (e.g.,~[\onlinecite{Halas_dissociation_H2_TiO2}]), these are particularly sensitive to increasing temperature~\cite{Un-Sivan-sensitivity} (via the Arrhenius Law~(\ref{EQ:Arrhenius})). Conversely, low activation energy reactions (i.e., those which are typically efficient at room temperature) are those that may benefit much more from ``hot'' electrons; this was indeed demonstrated recently in~[\onlinecite{Boltasseva_LPR_2020}].

All the above implies, somewhat ironically, that ``hot'' electrons are (typically) useful only under conditions which are the {\em exact opposite} to those in which they were originally envisioned to be efficient! Whether there is a {\em practical} advantage for speeding up chemical reactions occurring across a large volume using such dilute suspensions of metal NPs remains to be proven.

{\bf Outlook. } Our re-interpretation~\cite{anti-Halas-Science-paper,Y2-eppur-si-riscalda}, as well as similar work by others~\cite{Baffou-Quidant-Baldi,Khurgin-Faraday-hot-es,Khurgin-nanophotonics-2020,Khurgin-Levy-ACS-Photonics-2020,Jain_viewpoint} shows that thermal effects are frequently important and even dominant factors in plasmon-assisted photocatalysis, such that their quantification is essential for the understanding of the mechanisms underlying every specific chemical reaction. However, while the simplistic thermal theory~(\ref{EQ:Arrhenius}) was sufficient to explain several ``high-profile'' studies, there are quite a few studies it cannot explain. When such a discrepancy is observed, the catalytic enhancement is usually ascribed to non-thermal carriers, see, for example,~[\onlinecite{Baldi-ACS-Nano-2018,Frontiera-2018,Liu-Everitt-Nano-research-2019,yu2019plasmonic,Giulia_2019,Boltasseva_LPR_2020}]
, to name just a few. However, the exponential sensitivity mentioned above implies that in order to rectify such conclusions, one has to improve the model's validity and to reach a quantitative match of the theory to the experiments.

First, the quantitative thermal theory has to be complemented with a quantitative model of chemical interactions, including electron transfer between the NPs and the chemical moieties as well as the reaction dynamics. This is essential for the isolation of thermal and non-thermal contributions when they co-exist. For instance, in~[\onlinecite{plasmonic_photocatalysis_Linic}], employing a Kinetic Isotope (KIE) measurement showed that the strong ($\sim 100\%$) thermal contribution is accompanied by a small ($\sim 1-3\%$) contribution which was interpreted as non-thermal action (although a thermal origin for this effect has been suggested, see discussion in~[\onlinecite{Y2-eppur-si-riscalda}]). Another example is low activation energy reactions in which the thermal channel is simultaneously relatively efficient and insensitive to further illumination~\cite{Boltasseva_LPR_2020}. A third example is reactions which exhibit both temperature-dependent reaction rates and photo-selectivity, which cannot be simply explained with Arrhenius theory \cite{Yugang_Sun,Halas-Nature-Energy-2020,Yugang_Sun,yu2018plasmonic,yu2019plasmonic,JACS_Xiong,Giulia_2019}. This extension of the theory will also allow one to take into consideration the type of chemical reaction; specifically, as so far we have re-analyzed studies involving only bond dissociation, it would be interesting to examine also reduction-oxidation reactions in which charge transfer is an integral part of the reaction.

Second, while our work discussed at length photo-catalysis experiments in hydrostatic and dry conditions, many of the experiments in the field involve fluid flow and electro-chemical charge transfer (in solution)~\cite{Tatsuma_2004,Willets_electrochemistry_2018,Caleb_Hill_Heating_Electrochemical_JPCC_2019,Cortes_Nano_lett_2019,Lange_2020,Lopez_electrochemistry_2020,Lu_electrochemistry_2020}. Understanding these experiments requires inclusion of various experimental elements such as heat convection and stirring~\cite{yu2019plasmonic}, reactant pressure and distribution within the catalyst bed~\cite{Liu-Everitt-Nano-Letters-2019}, the thermal properties of the electro-chemical cell, charge transfer properties (including the effects of voltage bias) etc.. First steps towards achieving this were accomplished in~[\onlinecite{Un-Sivan-sensitivity}] (where the effective medium thermal calculations of~[\onlinecite{Y2-eppur-si-riscalda}] were rectified vs. exact simulations, as a first step towards combining heat and fluid dynamics) and in~[\onlinecite{Caleb_Hill_Heating_Electrochemical_JPCC_2019}] (where a formulation combining heat and charge transfer was described).

Third, one has to account for the effect of high illumination intensities and high temperatures on the optical properties of the system. Indeed, in many cases, the ambient temperature is increased significantly, to increase the (dark) reaction yield, so that the material permittivities change significantly (see e.g.,~\cite{Shalaev_ellipsometry_gold,Shalaev_ellipsometry_silver,PT_Shen_ellipsometry_gold} for some recent high temperature ellipsometry studies of metals). Furthermore, in some experiments (e.g.,~[\onlinecite{Halas_Science_2018}]), the light-induced heating reached several and even many hundreds of degrees Kelvin, so that the reaction rate ceased to grow linearly (see detailed discussion in~[\onlinecite{thm_hot_e_faraday_discuss_2019}, p. 271], as well as in~[\onlinecite{Y2-eppur-si-riscalda}] and [\onlinecite{R2R}]). Since the thermal nonlinearity of the metal NPs (and even of the host dielectric) may be relatively strong (see~[\onlinecite{Sivan-Chu-high-T-nl-plasmonics,Gurwich-Sivan-CW-nlty-metal_NP,IWU-Sivan-CW-nlty-metal_NP}]), it it likely to be stronger than non-thermal effects.

Finally, the theory should include, when necessary, also considerations related with the non-locality of the electron response, as well as quantum mechanical considerations associated with electron state discretization~\cite{Govorov_1,hot_es_Atwater,GdA_hot_es,Khurgin_Landau_damping,Khurgin-Faraday-hot-es,Khurgin-Levy-ACS-Photonics-2020,Lischner_hot_es}, electron tunneling out of the metal NP~\cite{Khurgin-nanophotonics-2020,Giulia_Nat_Comm_2018} etc. all of which are essential for the quantification of the reaction process as a whole. One of the immediate applications of such theory would be a quantification of the efficiency of promising pathways such as the elongation of tunnelled carrier lifetimes - via hole scavengers~\cite{yu2019plasmonic}, dielectric cores (in the spirit of photodetection experiments)~\cite{Boltasseva_LPR_2020} or core-shell antenna-reactor structures~\cite{JACS_Xiong}.

Many of the above elements were already developed. However, a complete theory that includes {\em all} these components is yet to be compiled and implemented for a quantitative comparison to experimental data. This will allow us, as a community, to embark upon a proper re-evaluation of past results, on one hand, and to design meaningful experiments and interpret them thoroughly, on the other.

In parallel to the extension of the theoretical framework, the experimental characterization of the temperature profile has to be improved. Indeed, measuring the temperature at only two points (as e.g., in~[\onlinecite{Liu-Everitt-Nano-Letters-2019,Liu-Everitt-Nano-research-2019,Halas-Nature-Energy-2020}]) gives no information on {\em unavoidable}~\cite{Un-Sivan-sensitivity} transverse gradients. Such gradients may strongly affect reaction rates, yet are many times un-accounted for. High resolution thermometry techniques have been developed~\cite{Cahill_T_measure,Baumberg_SERS_T_measure,Pramod_Reddy_T_measure_2015,Lukin_T_measure,NV_centers_T_measurement_NTU,Orrit-Caldarola_T_measure,Orrit-Caldarola_T_measure_transient} and implemented in the context of photocatalysis~\cite{Cortes_Nano_lett_2019}, however, they are far from being sufficiently robust~\cite{Orrit-Caldarola_T_measure_transient}. Indeed, similar difficulties to measure the temperature distribution were observed also in other contemporary problems in nanophotonics, in particular, in perovskite films, see~[\onlinecite{perovskite_solar_cells,perovskite_solar_cells_comment}] and the ongoing exchange. 

While the above elements are necessary for the characterization of the macroscopic properties, many recent experiments addressed the challenging problem of measuring the reaction rate on the single NP level~\cite{Vadai_Dionne_hydrogenation,Gross_Nature_2017,Giessen_single_NP_hydrogenation,Cortes_Nano_lett_2019,Peng_Chen}. Beyond the obvious difficulty to obtain a measurable reaction rate from such small entities, this context poses several additional challenges such as the need to quantify distributions and currents created by gradients of local electromagnetic fields, temperature and charge, heat transfer on the particle level (including the Kapitza resistance) etc.. Furthermore, reactions in the presence of single nanoparticles are sensitive to specific shape, size and composition, which affects the catalytic properties of the nanoparticle; e.g., sharp edges and corners are known to contribute to the catalytic enhancement - with and without ``hot'' electrons~\cite{El_Sayed_review}.

{\bf Final Thoughts. } The comprehensive critical studies published over the last year or so (see~[\onlinecite{dyn_hot_e_faraday_discuss_2019,Khurgin-Faraday-hot-es,thm_hot_e_faraday_discuss_2019,anti-Halas-Science-paper,Y2-eppur-si-riscalda,R2R,Jain_viewpoint,Baffou-Quidant-Baldi,Khurgin-nanophotonics-2020}]) and the vibrant ongoing discussion in the field  seem to have re-directed the study of plasmon-assisted photocatalysis into a more reliable, quantitative scientific routine. Detailed thermal calculations are becoming more common and accessible (e.g.,~[\onlinecite{Baffou-Quidant-Baldi,Boltasseva_LPR_2020,yu2019plasmonic,Y2-eppur-si-riscalda,Un-Sivan-sensitivity}]) and the awareness to the characteristics of the temperature distribution~\cite{Y2-eppur-si-riscalda,Un-Sivan-sensitivity,Boltasseva_LPR_2020} and to proper temperature measurements has grown~\cite{Cahill_T_measure,metal_luminescence_Cahill_PNAS,Cortes_Nat_Comm_2017,Baldi-ACS-Nano-2018,Orrit-Caldarola_T_measure,Cortes_Nano_lett_2019,Liu-Everitt-Nano-research-2019,Orrit-Caldarola_T_measure_transient}. Advanced techniques are being developed and the use of fluorescence and Raman techniques is becoming more abundant in the context of plasmon-assisted photocatalysis~\cite{Majima_1,Majima_1,Baumberg_SERS_T_measure,Baumberg-Faraday,Peng_Chen,Frontiera-2018,Orrit-Caldarola_T_measure_transient}. Indeed, several convincing demonstrations of ``hot'' electron-driven reactions with proper thermal control have been reported~\cite{Baldi-ACS-Nano-2018, yu2018plasmonic,yu2019plasmonic,Boltasseva_LPR_2020}.

Yet, some problematic practices still persist, for instance, the use of different samples for the photocatalysis experiment and the thermocatalysis control~\cite{Halas-Nature-Energy-2020}, incorrect normalization by catalyst mass~\cite{Halas-Nature-Energy-2020} (see detailed discussion in~[\onlinecite{R2R}]) and wrong extraction of ``hot'' electron contribution to the reaction rates~\cite{Halas-Nature-Catalysis-2020}. Additional common misconceptions (such as the incorrect claim on the dominance of non-thermal effects over thermal effects at low illumination intensity~\cite{Govorov_ACS_phot_2017}, the absence of transverse temperature uniformities, the dependence of the number of ``hot'' electrons on particle size and shape (see correction of~\cite{Govorov_ACS_phot_2017} in~\cite{Dubi-Sivan-Faraday}) or the ability of gas flow to eliminate temperature gradients) still need to be corrected.

We wish to end this perspective with a more general note. We believe that the ongoing debate reviewed above proves that there is room within the current publishing climate for more critical {\em (self-)reflective} research of the type that we described in our recent manuscripts; this may even be considered as a sort of necessary gadfly. Thus, addressing the scientific criticism in one’s own papers, encouraging debate and 
publishing controversial (and sometimes editorially inconvenient) papers, will catalyse uprooting of incorrect claims, unfounded beliefs and wrong practices.

\bigskip

{\bf Erratum.} Following lengthy discussions with L. Besteiro, we realized that one aspect of the criticism described in Ref.~[\onlinecite{Dubi-Sivan,Dubi-Sivan-APL-Perspective}] over his previous work~\cite{Govorov_ACS_phot_2017} is unjustified.

In particular, the formulation developed in Ref.~[\onlinecite{Govorov_ACS_phot_2017}] relies on a unique parameter defined as $T_{eff}$. While this parameter has no analog in similar formulations (see e.g., Refs.~[\onlinecite{hot_es_Atwater,GdA_hot_es,Dubi-Sivan})], and its origin is not clearly explained, its description can be easily lead one to understand that it represents the electron temperature. However, Besteiro clarified that this parameter is extracted from the numerical evaluation of the electron excitation/relaxation rate (Fig. 7(b)-(c) in Ref.~[\onlinecite{Govorov_ACS_phot_2017}]), such that despite its suggestive notation, it does not correspond to the electron temperature.

In that sense, our comparison of the choices of $T_{eff}$ to the predictions of classical heat equations (in Refs.~[\onlinecite{Dubi-Sivan,Dubi-Sivan-APL-Perspective}]) is irrelevant. We apologize for the confusion caused.

This issue, however, has only a small impact on the overall criticism expressed on the formulation of Ref.~[\onlinecite{Govorov_ACS_phot_2017}] in Refs.~[\onlinecite{Dubi-Sivan,Dubi-Sivan-Faraday,Dubi-Sivan-APL-Perspective}]. In a nutshell, that formulation assumes that the thermal and non-thermal distributions can be decoupled, and treats the latter only. As shown in Ref.~[\onlinecite{Dubi-Sivan,Dubi-Sivan-Faraday}], this decoupling is not valid when going beyond the relaxation time approximation for the electron-phonon interaction (as adopted in Ref.~[\onlinecite{Govorov_ACS_phot_2017}]) such that global energy conservation in the whole system (consisting of the photons, electrons, phonons and the environment) is considered. A potential exception is the case of very weak illumination. This subtle yet fundamental issue will be clarified in a future publication in detail.

\bigskip

{\bf Data Availability Statement} Data sharing is not applicable to this article as no new data were created or analyzed in this study.


\begin{thebibliography}{100}

\bibitem{Giannini_chemrev}
V.~Giannini, A.~I. Fern\'andez-Dom\'inguez, S.~C. Heck, and S.~A. Maier,
\newblock Chem. Rev. {\bf 111}, 3888 (2011).

\bibitem{Khurgin_Boltasseva_MRS}
J.~B. Khurgin and A.~Boltasseva,
\newblock MRS Bulletin {\bf 37}, 768 (2012).

\bibitem{Khurgin_loss_Nat_nanotech}
J.~B. Khurgin,
\newblock Nature Nanotechnology {\bf 10}, 2 (2015).

\bibitem{Dubi-Sivan}
Y.~Dubi and Y.~Sivan,
\newblock Light: Science and Applications: Nature {\bf 8}, 89 (2019).

\bibitem{Dubi-Sivan-Faraday}
Y.~Sivan, I.~W. Un, and Y.~Dubi,
\newblock Faraday Discussions {\bf 214}, 215 (2019).

\bibitem{thermo-plasmonics-review}
G.~Baffou and R.~Quidant,
\newblock Laser Photon. Rev. {\bf 7}, 171 (2013).

\bibitem{Baffou_solvothermal}
H.~M.~L. Robert, F.~Kundrat, E.~B.-U. na, H.~Rigneault, S.~Monneret,
  R.~Quidant, J.~Polleux, and G.~Baffou,
\newblock ACS Omega {\bf 1}, 2 (2016).

\bibitem{refractory_plasmonics}
U.~Guler, A.~Boltasseva, and V.~M. Shalaev,
\newblock Science {\bf 334}, 263 (2014).

\bibitem{solar_steam_apps}
Y.~Lin, H.~Xu, X.~Shan, Y.~Di, A.~Zhao, Y.~Hua, and Z.~Gan,
\newblock J. Mater. Chem. A {\bf 7}, 19203 (2019).

\bibitem{solar_steam_apps_2}
M.~Gao, L.~Zhu, C.~K. Peh, and G.~W. Ho,
\newblock Energy Environ. Sci. {\bf 12}, 841 (2019).

\bibitem{wei2018quantum}
Q.~Wei, S.~Wu, and Y.~Sun,
\newblock Advanced Materials {\bf 30}, 1802082 (2018).

\bibitem{zhang2016fundamental}
Y.~Zhang, C.~Yam, and G.~C. Schatz,
\newblock The Journal of Physical Chemistry Letters {\bf 7}, 1852 (2016).

\bibitem{seemala2019plasmon}
B.~Seemala, A.~J. Therrien, M.~Lou, K.~Li, J.~P. Finzel, J.~Qi, P.~Nordlander,
  and P.~Christopher,
\newblock ACS Energy Letters {\bf 4}, 1803 (2019).

\bibitem{Halas_Science_2018}
L.~Zhou, D.~F. Swearer, C.~Zhang, H.~Robatjazi, H.~Zhao, L.~Henderson, L.~Dong,
  P.~Christopher, E.~A. Carter, P.~Nordlander, and N.~J. Halas,
\newblock Science {\bf 362}, 69 (2018).

\bibitem{russell2003room}
K.~Russell, I.~Appelbaum, H.~Temkin, C.~Perry, V.~Narayanamurti, M.~Hanson, and
  A.~Gossard,
\newblock Applied physics letters {\bf 82}, 2960 (2003).

\bibitem{sandhu2000near}
J.~Sandhu, A.~Heberle, B.~Alphenaar, and J.~Cleaver,
\newblock Applied Physics Letters {\bf 76}, 1507 (2000).

\bibitem{Moskovits_hot_es_water_splitting}
J.~Lee, S.~Mubeen, X.~Ji, G.~Stucky, and M.~Moskovits,
\newblock Nano Lett. {\bf 12}, 5014 (2012).

\bibitem{Moskovits_photosensitization}
S.~Mubeen, G.~Hernandez-Sosa, D.~Moses, J.~Lee, and M.~Moskovits,
\newblock Nano Lett. {\bf 11}, 5548 (2011).

\bibitem{Uriel_Schottky}
I.~Goykhman, B.~Desiatov, J.~Khurgin, J.~Shappir, and U.~Levy,
\newblock Nano Lett. {\bf 11}, 2219 (2011).

\bibitem{Uriel_Schottky2}
I.~Goykhman, B.~Desiatov, J.~Khurgin, J.~Shappir, and U.~Levy,
\newblock Opt. Exp. {\bf 20}, 28594 (2012).

\bibitem{Valentine_hot_e_review}
W.~Li and J.~Valentine,
\newblock Nanophotonics {\bf 6}, 177 (2016).

\bibitem{Giulia_Nat_Comm_2018}
G.~Tagliabue, A.~S. Jermyn, R.~Sundararaman, A.~J. Welch, J.~S. DuChene,
  R.~Pala, A.~R. Davoyan, P.~Narang, and H.~A. Atwater,
\newblock Nature Communications {\bf 9}, 3394 (2018).

\bibitem{Halas_dissociation_H2_TiO2}
S.~Mukherjee, F.~Libisch, N.~Large, O.~Neumann, L.~V. Brown, J.~Cheng, J.~B.
  Lassiter, E.~A. Carter, P.~Nordlander, and N.~J. Halas,
\newblock Nano Lett. {\bf 13}, 240 (2013).

\bibitem{Halas_H2_dissociation_SiO2}
S.~Mukherjee, L.~Zhou, A.~Goodman, N.~Large, C.~Ayala-Orozco, Y.~Zhang,
  P.~Nordlander, and N.~J. Halas,
\newblock J. Am. Chem. Soc. {\bf 136}, 64 (2014).

\bibitem{christopher2011visible}
P.~Christopher, H.~Xin, and S.~Linic,
\newblock Nature Chemistry {\bf 3}, 467 (2011).

\bibitem{plasmonic_photocatalysis_Linic}
P.~Christopher, H.~Xin, A.~Marimuthu, and S.~Linic,
\newblock Nat. Materials {\bf 11}, 1044 (2012).

\bibitem{liu2018metal}
L.~Liu and A.~Corma,
\newblock Chem. Rev. {\bf 118}, 4981 (2018).

\bibitem{chem_rev_photochemistry_2006}
K.~Watanabe, D.~Menzel, N.~Nilius, and H.-J. Freund,
\newblock Chem. Rev. {\bf 106}, 4301 (2006).

\bibitem{thermal_shortening_catalyst_lifetime}
C.~T. Campbell, S.~C. Parker, and D.~E. Starr,
\newblock Science {\bf 298}, 811–814 (2002).

\bibitem{hot_e_review_Purdue}
A.~Naldoni, F.~Riboni, U.~Guler, A.~Boltasseva, V.~M. Shalaev, and A.~V.
  Kildishev,
\newblock Nanophotonics {\bf 5}, 112 (2016).

\bibitem{hirakawa2004photoinduced}
T.~Hirakawa and P.~V. Kamat,
\newblock Langmuir {\bf 20}, 5645 (2004).

\bibitem{plasmonic-chemistry-Baffou}
G.~Baffou and R.~Quidant,
\newblock Chem. Soc. Rev. {\bf 43}, 3898 (2014).

\bibitem{plasmonic_photocatalysis_Clavero}
C.~Clavero,
\newblock Nat. Photon. {\bf 8}, 95 (2014).

\bibitem{hot_es_review_2015_Moskovits}
M.~Moskovits,
\newblock Nature Nanotech. {\bf 10}, 6 (2015).

\bibitem{hou2013review}
W.~Hou and S.~B. Cronin,
\newblock Advanced Functional Materials {\bf 23}, 1612 (2013).

\bibitem{wu2017direct}
X.~Wu, E.~Jaatinen, S.~Sarina, and H.~Y. Zhu,
\newblock Journal of Physics D {\bf 50}, 283001 (2017).

\bibitem{brooks2018toward}
J.~L. Brooks, C.~L. Warkentin, D.~Saha, E.~L. Keller, and R.~R. Frontiera,
\newblock Nanophotonics {\bf 7}, 1697 (2018).

\bibitem{zhang2017surface}
Y.~Zhang, S.~He, W.~Guo, Y.~Hu, J.~Huang, J.~R. Mulcahy, and W.~D. Wei,
\newblock Chemical reviews {\bf 118}, 2927 (2017).

\bibitem{Wolf-Ertl}
M.~Bonn, S.~Funk, C.~Hess, D.~N. Denzler, C.~Stampfl, M.~Scheffler, M.~Wolf,
  and G.~Ertl,
\newblock Science {\bf 285}, 1042 (1999).

\bibitem{Petek_Nat_Phot_2017}
S.~Tan, A.~Argondizzo, J.~Ren, L.~Liu, J.~Zhao, and H.~Petek,
\newblock Nature Photonics {\bf 11}, 806–812 (2017).

\bibitem{Y2-eppur-si-riscalda}
Y.~Sivan, I.~W. Un, and Y.~Dubi,
\newblock Chemical Sciences {\bf 11}, 5017 (2020).

\bibitem{Naldoni-tutorial-2020}
L.~Mascaretti and A.~Naldoni,
\newblock J. Appl. Phys. {\bf 128}, 041101 (2020).

\bibitem{Aeschliman_e_photoemission_review}
M.~Bauer, A.~Marienfeld, and M.~Aeschlimann,
\newblock Progress in Surface Science {\bf 90}, 319–376 (2015).

\bibitem{Liu-Everitt-Nano-Letters-2019}
X.~Li, X.~Zhang, H.~O. Everitt, and J.~Liu,
\newblock Nano Letters {\bf 19}, 1706 (2019).

\bibitem{JACS_Xiong}
H.~Huang, L.~Zhang, Z.~Lu, R.~Long, C.~Zhang, Y.~Lin, K.~Wei, C.~Wang, L.~Chen,
  Z.-Y. Li, Q.~Zhang, Y.~Luo, and Y.~Xiong,
\newblock J. Am. Chem. Soc. {\bf 138}, 6822 (2016).

\bibitem{Swearer_antenna_reactor}
F.~Swearer, H.~Zhao, L.~Zhou, C.~Zhang, H.~Robatjazi, J.~M.~P. Martirez, C.~M.
  Krauter, S.~Yazdi, M.~J. McClain, E.~Ringe, E.~A. Carter, P.~Nordlander, and
  N.~J. Halas,
\newblock Proc. Natl. Acad. Sci. U.S.A {\bf 113}, 8916 (2016).

\bibitem{Boltasseva_LPR_2020}
X.~Xu, A.~Dutta, J.~Khurgin, A.~Wei, V.~M. Shalaev, and A.~Boltasseva,
\newblock Laser and Photonics Reviews {\bf 14}, 1900376 (2020).

\bibitem{culver1996temperature}
J.~Culver, M.~Li, Z.-J. Sun, R.~Hochstrasser, and A.~Yodh,
\newblock Chemical physics {\bf 205}, 159 (1996).

\bibitem{leenheer2014solar}
A.~J. Leenheer, P.~Narang, N.~S. Lewis, and H.~A. Atwater,
\newblock Journal of Applied Physics {\bf 115}, 134301 (2014).

\bibitem{Baldi-ACS-Nano-2018}
R.~Kamarudheen, G.~W. Castellanos, L.~P.~J. Kamp, H.~J.~H. Clercx, and
  A.~Baldi,
\newblock ACS Nano {\bf 12}, 8447 (2018).

\bibitem{sarhan2019importance}
R.~M. Sarhan, W.~Koopman, R.~Schuetz, T.~Schmid, F.~Liebig, J.~Koetz, and
  M.~Bargheer,
\newblock Scientific reports {\bf 9}, 3060 (2019).

\bibitem{Govorov_thermoplasmonics}
A.~O. Govorov and H.~H. Richardson,
\newblock Nano Today {\bf 2}, 30 (2007).

\bibitem{anti-Halas-Science-paper}
Y.~Sivan, J.~Baraban, I.~W. Un, and Y.~Dubi,
\newblock Science {\bf 364}, eaaw9367 (2019).

\bibitem{R2R}
Y.~Sivan, J.~Baraban, and Y.~Dubi,
\newblock OSA Continuum {\bf 3}, 483 (2020).

\bibitem{Un-Sivan-sensitivity}
I.~W. Un and Y.~Sivan,
\newblock Nanoscale, accepted; https://arxiv.org/abs/2007.03421  (2020).

\bibitem{thm_hot_e_faraday_discuss_2019}
J.~Aizpurua, F.~Baletto, J.~Baumberg, P.~Christopher, B.~d. Nijs, P.~Deshpande,
  Y.~Diaz~Fernandez, L.~Fabris, S.~Freakley, S.~Gawinkowski, A.~Govorov,
  N.~Halas, R.~Hernandez, B.~Jankiewicz, J.~Khurgin, M.~Kuisma, P.~V. Kumar,
  J.~Lischner, J.~Liu, A.~Marini, R.~J. Maurer, N.~S. Mueller, M.~Parente,
  J.~Y. Park, S.~Reich, Y.~Sivan, G.~Tagliabue, L.~Torrente-Murciano,
  M.~Thangamuthu, X.~Xiao, and A.~Zayats,
\newblock Faraday Discussions {\bf 214}, 245 (2019).

\bibitem{dyn_hot_e_faraday_discuss_2019}
J.~Aizpurua, M.~Ashfold, F.~Baletto, J.~Baumberg, P.~Christopher, E.~Cortes,
  B.~de~Nijs, Y.~Diaz~Fernandez, J.~Gargiulo, S.~Gawinkowski, N.~Halas,
  R.~Hamans, B.~Jankiewicz, J.~Khurgin, P.~V. Kumar, J.~Liu, S.~Maier, R.~J.
  Maurer, A.~Mount, N.~S. Mueller, R.~Oulton, M.~Parente, J.~Y. Park,
  J.~Polanyi, J.~Quiroz, S.~Rejman, S.~Schlucker, Z.~Schultz, Y.~Sivan,
  G.~Tagliabue, M.~Thangamuthu, L.~Torrente-Murciano, X.~Xiao, A.~Zayats, and
  C.~Zhan,
\newblock Faraday Discussions {\bf 214}, 123 (2019).

\bibitem{Baffou-Quidant-Baldi}
G.~Baffou, I.~Bordacchini, A.~Baldi, and R.~Quidant,
\newblock {Light: Science and Applications} {\bf 9}, 108 (2020).

\bibitem{Khurgin-Faraday-hot-es}
J.~Khurgin,
\newblock Faraday Discussions {\bf 214}, 35 (2019).

\bibitem{Khurgin-nanophotonics-2020}
J.~B. Khurgin,
\newblock Nanophotonics {\bf 9}, 453 (2020).

\bibitem{Jain_viewpoint}
P.~K. Jain,
\newblock J. Phys. Chem. {\bf 123}, 24347 (2019).

\bibitem{Manjavacas_Nordlander}
A.~Manjavacas, J.~G. Liu, V.~Kulkarni, and P.~Nordlander,
\newblock ACS Nano {\bf 8}, 7630 (2014).

\bibitem{Govorov_1}
A.~O. Govorov, H.~Zhang, and Y.~K. Gun'ko,
\newblock J. Phys. Chem. C {\bf 117}, 16616 (2013).

\bibitem{Govorov_ACS_phot_2017}
L.~V. Besteiro, X.-T. Kong, Z.~Wang, G.~Hartland, and A.~O. Govorov,
\newblock ACS Photonics {\bf 4}, 2759 (2017).

\bibitem{Note1}
Indeed, in~\cite {Y2-eppur-si-riscalda} we have shown that $T_M$ was several
  tens of degrees lower than the actual temperature, and sometimes~(\cite
  {Halas_Science_2018,R2R}) even several hundreds of degrees lower.

\bibitem{thermo-plasmonics-basics}
G.~Baffou, R.~Quidant, and F.~J.~G. de~Abajo,
\newblock ACS Nano {\bf 4}, 709 (2010).

\bibitem{Baffou_pulsed_heat_eq_with_Kapitza}
G.~Baffou and H.~Rigneault,
\newblock Phys. Rev. B {\bf 84}, 035415 (2011).

\bibitem{thermo-plasmonics-multi_NP}
G.~Baffou, P.~Berto, E.~B. Urena, R.~Quidant, S.~Monneret, J.~Polleux, and
  H.~Rigneault,
\newblock ACS Nano {\bf 7}, 6478 (2013).

\bibitem{jain2019phenomenological}
P.~K. Jain,
\newblock Chem. Sci  (2020).

\bibitem{response_2_Jain}
Y.~Dubi, I.~W. Un, and Y.~Sivan,
\newblock Chem. Sci  (2020).

\bibitem{Liu-Everitt-Nano-research-2019}
X.~Li, H.~O. Everitt, and J.~Liu,
\newblock Nano Research {\bf 19}, 1706 (2019).

\bibitem{Halas-Nature-Catalysis-2020}
H.~Robatjazi, J.~L. Bao, M.~Zhang, L.~Zhou, P.~Christopher, E.~A. Carter,
  P.~Nordlander, and N.~J. Halas,
\newblock Nature Catalysis {\bf 3}, 564 (2020).

\bibitem{Note2}
Note that the transverse gradients were not accounted for in that work.

\bibitem{yu2019plasmonic}
S.~Yu and P.~K. Jain,
\newblock Nature Communications {\bf 10}, 2022 (2019).

\bibitem{Khurgin-Levy-ACS-Photonics-2020}
J.~B. Khurgin and U.~Levy,
\newblock ACS Photonics {\bf 7}, 547 (2020).

\bibitem{Frontiera-2018}
E.~L. Keller and R.~R. Frontiera,
\newblock ACS Nano {\bf 12}, 5848 (2018).

\bibitem{Giulia_2019}
A.~J. Welch, J.~S. DuChene, G.~Tagliabue, A.~Davoyan, W.-H. Cheng, and H.~A.
  Atwater,
\newblock ACS Appl. Energy Mater. {\bf 2}, 164 (2019).

\bibitem{Yugang_Sun}
X.~Dai, Q.~Wei, T.~Duong, and Y.~Sun,
\newblock Chem. Nano. Mat. {\bf 5}, 1000 (2019).

\bibitem{Halas-Nature-Energy-2020}
L.~Zhou, J.~M.~P. Martirez, J.~Finzel, C.~Zhang, D.~F. Swearer, S.~Tian,
  H.~Robatjazi, M.~Lou, L.~Dong, L.~Henderson, P.~Christopher, E.~A. Carter,
  P.~Nordlander, and N.~J. Halas,
\newblock Nature Energy {\bf 5}, 61 (2020).

\bibitem{yu2018plasmonic}
S.~Yu, A.~J. Wilson, J.~Heo, and P.~K. Jain,
\newblock Nano letters {\bf 18}, 2189 (2018).

\bibitem{Tatsuma_2004}
Y.~Tian and T.~Tatsuma,
\newblock Chem. Commun. , 1810 (2004).

\bibitem{Willets_electrochemistry_2018}
Y.~Yu, V.~Sundaresan, and K.~A. Willets,
\newblock J. Phys. Chem. C {\bf 122}, 5040 (2018).

\bibitem{Caleb_Hill_Heating_Electrochemical_JPCC_2019}
M.~Maley, J.~W. Hill, P.~Saha, J.~D. Walmsley, and C.~M. Hill,
\newblock The Journal of Physical Chemistry C {\bf 123}, 12390 (2019).

\bibitem{Cortes_Nano_lett_2019}
E.~Pensa, J.~Gargiulo, A.~Lauri, S.~Schl\"ucker, E.~Cort\'es, and S.~A. Maier,
\newblock Nano Lett. {\bf 19}, 1867 (2019).

\bibitem{Lange_2020}
M.~Rodio, M.~Graf, F.~Schulz, N.~S. Mueller, M.~Eich, and H.~Lange,
\newblock ACS Catalysis {\bf 10}, 2345 (2020).

\bibitem{Lopez_electrochemistry_2020}
N.~B. Schorr, M.~J. Counihan, R.~Bhargava, and J.~Rodr\'iguez-Lo\'pez,
\newblock Anal. Chem. {\bf 92}, 3666 (2020).

\bibitem{Lu_electrochemistry_2020}
W.~Ou, B.~Zhou, J.~Shen, T.~W. Lo, D.~Lei, S.~Li, J.~Zhong, Y.~Y. Li, and
  J.~Lu,
\newblock Angew. Chem. Int. Ed. {\bf 59}, 6790 (2020).

\bibitem{Shalaev_ellipsometry_gold}
H.~Reddy, U.~Guler, A.~V. Kildishev, A.~Boltasseva, and V.~M. Shalaev,
\newblock Optical Materials Express {\bf 6}, 2776 (2016).

\bibitem{Shalaev_ellipsometry_silver}
H.~Reddy, U.~Guler, K.~Chaudhuri, A.~Dutta, A.~V. Kildishev, V.~M. Shalaev, and
  A.~Boltasseva,
\newblock ACS Photonics {\bf 4}, 1083 (2017).

\bibitem{PT_Shen_ellipsometry_gold}
P.-T. Shen, Y.~Sivan, C.-W. Lin, H.-L. Liu, C.-W. Chang, and S.-W. Chu,
\newblock Opt. Exp. {\bf 24}, 19254 (2016).

\bibitem{Sivan-Chu-high-T-nl-plasmonics}
Y.~Sivan and S.-W. Chu,
\newblock Nanophotonics {\bf 6}, 317 (2017).

\bibitem{Gurwich-Sivan-CW-nlty-metal_NP}
I.~Gurwich and Y.~Sivan,
\newblock Phys. Rev. E {\bf 96}, 012212 (2017).

\bibitem{IWU-Sivan-CW-nlty-metal_NP}
I.~W. Un and Y.~Sivan,
\newblock https://arxiv.org/abs/2006.11884 .

\bibitem{hot_es_Atwater}
A.~M. Brown, R.~Sundararaman, P.~Narang, W.~A. Goddard, and H.~A. Atwater,
\newblock ACS Nano {\bf 10}, 957 (2016).

\bibitem{GdA_hot_es}
J.~R.~M. Saavedra, A.~Asenjo-Garcia, and F.~J.~G. de~Abajo,
\newblock ACS Photonics {\bf 3}, 1637 (2016).

\bibitem{Khurgin_Landau_damping}
J.~Khurgin, W.-Y. Tsai, D.~P. Tsai, and G.~Sun,
\newblock ACS Photonics {\bf 4}, 2871 (2017).

\bibitem{Lischner_hot_es}
S.~D. Forno, L.~Ranno, and J.~Lischner,
\newblock J. Phys. Chem. C. {\bf 122}, 8517 (2018).

\bibitem{Cahill_T_measure}
X.~Xie and D.~G. Cahill,
\newblock Appl. Phys. Lett. {\bf 109}, 183104 (2016).

\bibitem{Baumberg_SERS_T_measure}
J.~T. Hugall and J.~J. Baumberg,
\newblock Nano Lett. {\bf 15}, 2600 (2015).

\bibitem{Pramod_Reddy_T_measure_2015}
K.~Kim, B.~Song, V.~Fern\'andez-Hurtado, W.~Lee, W.~Jeong, L.~Cui, D.~Thompson,
  J.~Feist, M.~T.~H. Reid, F.~J. Garc\'ia-Vidal, J.~C. Cuevas, E.~Meyhofer, and
  P.~Reddy,
\newblock Nature {\bf 528}, 387 (2015).

\bibitem{Lukin_T_measure}
G.~Kucsko, P.~C. Maurer, N.~Y. Yao, M.~Kubo, H.~J. Noh, P.~K. Lo, H.~Park, and
  M.~D. Lukin,
\newblock Nature {\bf 500}, 54 (2013).

\bibitem{NV_centers_T_measurement_NTU}
Y.-K. Tzeng, P.-C. Tsai, H.-Y. Liu, O.~Chen, H.~Hsu, F.-G. Yee, M.-S. Chang,
  and H.-C. Chang,
\newblock Nano Lett. {\bf 15}, 3945 (2015).

\bibitem{Orrit-Caldarola_T_measure}
A.~Carattino, M.~Caldarola, and M.~Orrit,
\newblock Nano Lett. {\bf 18}, 874 (2017).

\bibitem{Orrit-Caldarola_T_measure_transient}
T.~Jollans, M.~Caldarola, Y.~Sivan, and M.~Orrit,
\newblock Journal of Physical Chemistry A, accepted;
  https://arxiv.org/abs/2003.08790  (2020).

\bibitem{perovskite_solar_cells}
H.~Tsai, R.~Asadpour, J.-C. Blancon, C.~C. Stoumpos, O.~Durand, J.~W. Strzalka,
  B.~Chen, R.~Verduzco, P.~M. Ajayan, S.~Tretiak, J.~Even, M.~A. Alam, M.~G.
  Kanatzidis, W.~Nie, and A.~D. Mohite,
\newblock Science {\bf 360}, 67 (2018).

\bibitem{perovskite_solar_cells_comment}
N.~Rolston, R.~Bennett-Kennett, L.~T. Schelhas, J.~M. Luther, J.~A. Christians,
  J.~J. Berry, and R.~H. Dauskardt,
\newblock Science {\bf 368}, eaay8691 (2020).

\bibitem{Vadai_Dionne_hydrogenation}
M.~Vadai, D.~Angell, F.~Hayee, K.~Sywtu, and J.~Dionne,
\newblock Nature Communications {\bf 9}, 4658 (2018).

\bibitem{Gross_Nature_2017}
C.-Y. Wu, W.~J. Wolf, Y.~Levartovsky, H.~A. Bechtel, M.~C. Martin, F.~D. Toste,
  and E.~Gross,
\newblock Nature {\bf 541}, 511 (2017).

\bibitem{Giessen_single_NP_hydrogenation}
J.~Karst, F.~Sterl, H.~Linnenbank, T.~Weiss, M.~Hentschel, and H.~Giessen,
\newblock Science Advances {\bf 6}, eaaz0566 (2020).

\bibitem{Peng_Chen}
N.~Zou, G.~Chen, X.~Mao, H.~Shen, E.~Choudhary, X.~Zhou, and P.~Chen,
\newblock Nano Lett. {\bf 12}, 5570 (2018).

\bibitem{El_Sayed_review}
M.~B. Mohamed, R.~Narayanan, and M.~A. El-Sayed,
\newblock Acc. Chem. Res. {\bf 46}, 1795 (2013).

\bibitem{metal_luminescence_Cahill_PNAS}
J.~Huang, W.~Wang, C.~J. Murphy, and D.~G. Cahill,
\newblock Proc. Nat. Acad. Sci. U.S.A {\bf 111}, 906 (2014).

\bibitem{Cortes_Nat_Comm_2017}
E.~Cort\'es, W.~Xie, J.~Cambiasso, A.~S. Jermyn, R.~Sundararaman, P.~Narang,
  S.~Schl\"ucker, and S.~A. Maier1,
\newblock Nature Communications {\bf 8}, 14880 (2017).

\bibitem{Majima_1}
Z.~Zheng, T.~Tachikawa, and T.~Majima,
\newblock J. Am. Chem. Soc. {\bf 136}, 6870 (2014).

\bibitem{Baumberg-Faraday}
J.~Huang, B.~de~Nijs, S.~Cormier, K.~Sokolowski, D.-B. Grys, C.~A. Readman,
  S.~J. Barrow, O.~A. Schermanb, and J.~J. Baumberg,
\newblock Faraday Discussions {\bf 214}, 455 (2019).

\bibitem{Dubi-Sivan-APL-Perspective}
Y.~Sivan and Y.~Dubi,
\newblock Appl. Phys. Letters {\bf 117}, 130501 (2020).

\end{thebibliography}

\end{document}